\documentstyle[aps,prl,psfig]{revtex}

\begin{document}

\title{ Effects of Neutron Spatial Distributions on Atomic Parity
Nonconservation in Cesium. }

\author{S.~J.~Pollock
\thanks{e-mail: Steven.Pollock@Colorado.edu}
and M.~C.~Welliver
\thanks{e-mail: welliver@lucky.colorado.edu} }
\address{Dep't of Physics, University of Colorado, Boulder CO 80309 }

\preprint{CU Preprint NPL-1163}
\date{July 12, 1999}
\maketitle


\begin{abstract}

We have examined modifications to the nuclear weak charge due to small
differences between the spatial distributions of neutrons and protons
in the Cs nucleus.  We derive approximate formulae to estimate the
value and uncertainty of this modification based only on nuclear rms
neutron and proton radii. Present uncertainties in neutron
distributions in Cs are difficult to quantify, but we conclude that
they should not be neglected when using atomic parity nonconservation
experiments as a means to test the Standard Model.

\end{abstract}


Recent measurements of transition polarizabilities~\cite{Wieman99},
coupled with previous measurements of parity nonconservation (PNC) in
atomic cesium~\cite{Wieman98} have significantly reduced uncertainties
associated with the extraction of $Q_{{w}}$, the radiatively
corrected weak charge of the Cs nucleus. The latest result\cite{Wieman99},
$Q_{{w}}^{\rm expt} = -72.06(28)_{\rm expt}\ (34)_{\rm atomic\
theory}$ is in mild disagreement, at the $2.5\sigma$
level, with the Standard Model prediction of $Q_{{w}}^{\rm
St. Mod.} = -73.20(13)_{\rm theory}$.~\cite{Marciano} The experimental
number requires input from atomic theory
calculations~\cite{Johnson,Dzuba} which include effects of
normalization of the relevant axial electron transition matrix element
in the vicinity of the nucleus. The finite nuclear size is
incorporated by including $\rho_{{N}}(r)$, the spatial nuclear
distribution, in the matrix elements. One possible contribution to
$Q_{{w}}$ which has been left out of the quoted numbers is the
modification of the extracted weak charge due to the difference
between neutron and proton spatial distributions in this nucleus with
relatively large neutron excess.

The effect of the neutron distribution differing from the proton
distribution in a nucleus has been explicitly considered in the atomic
theory calculations,~\cite{Johnson} and was dismissed because the
estimated size was extremely small compared to existing uncertainties
at the time.  Other authors~\cite{Fortson,Chen} have
also derived and discussed this contribution further. In the case of
Cs, all authors agree the effect is quite small.  However, with the
significant reduction in errors in recent atomic PNC measurements,
the effect should no longer be neglected. As we argue below, the
additional uncertainties in extracting $Q_{{w}}$ from the data
arising from neutron-proton distribution differences are slightly
below the uncertainties arising from atomic theory calculations or
current experimental error bars, but are comparable to Standard Model
radiative correction uncertainties.

In this note, we attempt to quantify the additive contribution and
uncertainties to the nuclear weak charge, $Q_{{w}}$, arising
from the relatively poorly known spatial distribution of neutrons in
the nucleus, $\rho_{{n}}(r)$.  We briefly summarize some
relevant nuclear structure issues, both theoretical and experimental.
We also briefly discuss methods that could improve this knowledge. We
present results of our numerical calculations of $Q_{{w}}$
arising from various $\rho_{{n}}$ distributions, and present
approximate methods which show what effect differing nuclear structure
model predictions would have on precision Standard Model tests.

At tree level in the Standard Model, the nuclear weak charge is
$Q_{{w}}^{\rm St. Mod} = (1-4\sin^2\theta_W)Z - N$, with N and
Z the neutron and proton number, and $\sin^2\theta_W$ the weak mixing
angle. Standard Model radiative corrections modify this formula
slightly.~\cite{Marciano} The effect of finite nuclear extent is to
modify N and Z to $q_{{n}} N$ and $q_{{p}} Z$
respectively,~\cite{Fortson} where
\begin{equation}
q_{{n} ({p})} = \int f(r)
\rho_{{n} ({p})}(r) d^{3} r .
\label{weakcharge}
\end{equation}
Here $f(r)$ is a folding function determined from the radial
dependence of the electron axial transition matrix element inside the
nucleus, and the neutron (proton) spatial distribution
$\rho_{{n} ({p})}$ is normalized to unity.  It is common
to characterize the neutron distribution by its rms value,
$R_{{n}}$, since it can easily be shown that the weak charge is
most sensitive to this moment.  To the extent that $\rho_{{n}}$
and $\rho_{{p}}$ are the same, the overall nuclear size effect
can be completely factored out. This has explicitly been done in the
experimental extraction of $Q_{{w}}$. The slight difference
between $q_{{n}}$ and $q_{{p}}$ has the effect of
modifying the effective weak charge:
\begin{equation}
Q_{{w}}=Q_{{w}}^{\rm St. Mod} + \Delta
Q_{{w}}^{{n-p}},
\label{weakchargeii}
\end{equation}
where 
\footnote{There will be additional small multiplicative
corrections to $\Delta Q_{{w}}^{{n-p}}$ arising from
Standard Model radiative corrections, as well as additive corrections
arising from e.g. internal structure of the nucleon, but these can be
safely neglected since $\Delta Q_{{w}}^{{n-p}}$ is
itself so small.}
\begin{equation}
\Delta Q_{{w}}^{{n-p}} = N(1-q_{{n}}/q_{{p}}).
\label{chargeshift}
\end{equation}

A naive calculation,~\cite{Fortson} helpful for quick estimates of the
effect of different possible neutron distributions on $\Delta
Q_{{w}}^{{n-p}}$, can be made by assuming a uniform
nuclear charge distribution (zero-temperature Fermi gas), and then
parameterizing the neutron distribution solely by its value of
$R_{{n}}$. In this approximation, one solves the Dirac equation
for the electron axial matrix elements, $f(r)$, near the origin by
expanding in powers of $\alpha$ (the fine structure constant).
Finally, we can assume
$R_{{n}} \approx R_{{p}}$, characterizing the difference
by a single small parameter, $(R_{{n}}^2/ R_{{p}}^2)
\equiv 1+\epsilon$.  In this case, we
find~\cite{Fortson}
\begin{eqnarray}
q_{{p}} \approx & 1-(Z\alpha)^2(.26),\\
\label{approxi}
q_{{n}} \approx & 1-(Z\alpha)^2(.26+.221\epsilon),\\
\label{approxii}
\Delta Q_{{w}}^{{n-p}} \approx & N (Z\alpha)^2 (.221
\epsilon)/q_{{p}} .
\label{approxiii}
\end{eqnarray} 

\begin{figure}
\centerline{\psfig{figure=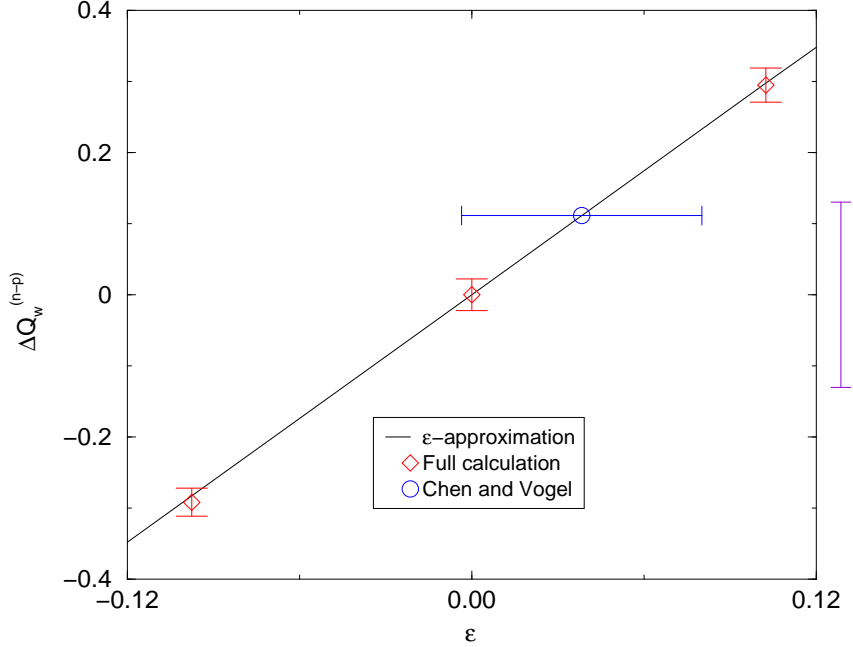,width=4.4in}}
\caption{Correction to the weak charge of Cs due to differences between
neutron and proton spatial distributions as a function of $\epsilon
\equiv R_{{n}}^{2} / R_{{p}}^{2} - 1$.  The solid curve is the
prediction of Eq.~\ref{approxiii}.  The diamonds are generated from a
full numerical calculation with a realistic charge distribution,
scaling $\rho_p$ to get $\rho_n$.  The error bars on the diamonds arise
by varying the surface thickness of $\rho_n$, keeping $\epsilon$
fixed.  The circle uses Eq.~\ref{approxiii} assuming the neutron radius
and uncertainty of ref.~\protect\cite{Chen}. The vertical error bar
to the side of the plot shows just the \emph{uncertainty} in the
Standard Model prediction~\protect\cite{Marciano} of $Q_W$.}
\label{cs_plot} \end{figure}

Eq.~\ref{approxii} shows the rough dependence of the correction to
the weak charge on the difference between neutron and proton
distributions, characterized by $\epsilon$.  Results of this naive
calculation are shown as the solid line in Fig.~\ref{cs_plot}.  The
slope of the line demonstrates the sensitivity of the uncertainty in
$\Delta Q_{{w}}^{{n-p}}$ to the uncertainty in rms
neutron radius.  The range in $\epsilon$ of $\pm 0.1$ corresponds to a
$\delta R_{{n}} / R_{{n}}$ of about $5\%$, which we argue
below might be a reasonable estimate of the uncertainty in neutron rms
radius.  We do not need to rely on these approximations; we have solved
the Dirac Equation numerically for s- and p- state electron wave
functions given the experimental charge distribution of Cs, evaluated
$f(r)$ numerically, and thus calculated $q_{{n}}$,
$q_{{p}}$, and $\Delta Q_{{w}}^{{n-p}}$ given
various model predictions for the neutron distribution.  The diamonds
in Fig.~\ref{cs_plot} correspond to the full calculation assuming 
neutron distributions with the same shape as the proton
distribution, scaled to give the particular values of $\epsilon$.  The
above approximations prove to be accurate, although the resulting
uncertainty in $\Delta Q_{{w}}^{{n-p}}$ is marginally
underestimated by only including the uncertainty in $R_{{n}}$.
The additional effects of neutron distribution shape variations could
slightly increase the uncertainty in the nuclear contribution to the
weak charge, as demonstrated by the error bars on the diamonds in
Fig.~\ref{cs_plot}. These error bars arise by assuming a 2 parameter
Fermi fit for the neutron distribution, $\rho_n(r) = 1/(1+e^{[(r-c)/z]})$,
and allowing the ``skin thickness'' parameter $z_n$ to vary by $\pm 0.1$,
keeping $\epsilon$ fixed. Such a variation is comparable to the
difference $z_n-z_p$ in various nuclear models\cite{dobiii}.  
(A more detailed
analysis of the effect of neutron shape on $Q_{{w}}$ will be
presented elsewhere~\cite{Welliver}.) From Fig 1, it is clear that
the uncertainty in the radius $R_n$ dominates the uncertainty in 
$\Delta Q_{w}^{n-p}$.

The effect of $\Delta Q_{{w}}^{{n-p}}$ was understood and
estimated in the atomic structure calculations~\cite{Johnson} by first
assuming $\rho_{{n}}(r)=\rho_{{p}}(r)$ (so
$q_{{n}}=q_{{p}}$ and $\Delta
Q_{{w}}^{{n-p}}=0$) and then recalculating with a
theoretical parameterization of the neutron density.~\cite{Brack} The
resulting $\Delta Q_{{w}}^{{n-p}}\approx 0.06$ was
extremely small, amounting to about 0.08$\%$ of the total weak charge,
and was thereafter ignored.  This neutron density was obtained by
scaling a variational extended spherical Thomas-Fermi calculation
using an effective parameterization of the nuclear Lagrangian (
``Skyrme SkM*''), which happened to yield a nuclear neutron rms radius
which differed by only 0.9$\%$ from the proton rms radius.
Eq.~\ref{approxiii} confirms the size of this shift, given only the rms
neutron and proton radii.  However, the assumed $\rho_{{n}}(r)$
distribution may not be an accurate representation of the correct
neutron distribution. There exists both theoretical and experimental
evidence that $R_{{n}}$ might differ from $R_{{p}}$ by
significantly more than 0.9$\%$, and thus from Eq.~\ref{approxiii},
$\Delta Q_{{w}}^{{n-p}}$ may be similarly
underestimated.

In a more recent theoretical analysis, Chen and Vogel~\cite{Chen}
considered two more sophisticated nuclear structure models. Both models
involved a Skyrme parameterized nuclear Lagrangian,~\cite{Skyrme}
computed in the spherical Hartree-Fock (HF) approximation. Such models
are quite successful in predicting a wide variety of nuclear
observables, including charge distributions, binding energies, bulk
properties, etc.  These two models (SkM* and SkIII) yielded
$R_{{n}}/R_{{p}}$ values of 1.022 and 1.016
respectively.  Using the average of these values in Eq.~\ref{approxiii}
we obtain $\Delta Q_{{w}}^{{n-p}} = +.11$, double the
estimate of Ref. \cite{Johnson}.  Using spatial distributions of
neutron and proton densities from even more recent nuclear structure
models~\cite{dobiii}, we have calculated the nuclear correction directly, 
rather than
using the approximation of Eq.~\ref{approxiii}. 
Using spherical Skyrme SLy4 distributions, we find
$\Delta Q_{{w}}^{{n-p}} = +.14$.  Similarly, using
(spherical) Gogny distributions, including blocking, we
find $\Delta Q_{{w}}^{{n-p}} = +.11$ (Eq.~\ref{approxiii}
for these two cases predicts +.15 and +.12 respectively). Relativistic
potentials~\cite{Relativistic} typically generate significantly larger
neutron radii (see discussion below), and thus would predict larger
$\Delta Q_{{w}}^{{n-p}}$, possibly by a factor of 2 or
more, based on calculations in nearby nuclei, but no $^{133}$Cs
distributions for such models have been published to date.  Note that
if $R_n>R_p$, then $\Delta Q_{{w}}^{{n-p}}$ is positive. The central
value of the most recent experiment~\cite{Wieman99} gives
$Q_{{w}}=-72.06$, compared to $Q_{{w}}^{\rm St.  Mod}=
-73.20$, so this nuclear correction is of the right sign to partially
explain the small discrepancy.  However, if one wanted to attribute the
difference entirely to nuclear physics effects, one would require
$R_{{n}} = (1.18\pm
.07) R_{{p}}$ (adding all atomic experimental and theoretical,
and
Standard Model theoretical errors in quadrature), which is
{\it significantly} out of the range of any theoretical or experimental
nuclear structure predictions.

The fundamental question regarding nuclear structure remains --- what
{\it uncertainty} should be associated with $\Delta
Q_{{w}}^{{n-p}}$?  Chen and Vogel~\cite{Chen} argued
that a reasonable uncertainty in their calculated neutron radius might
be $\delta R_{{n}}^2\approx \pm 1 {\rm fm}^2$. According to 
Eq.~\ref{approxiii}, this corresponds to an uncertainty $\delta \Delta
Q_{{w}}^{{n-p}} = \pm 0.13$. The estimate in
ref.~\cite{Chen} for the theoretical uncertainty in $Q_{{w}}$
for a single isotope was slightly larger, 0.25$\%$ of
$Q_{{w}}$, i.e.  $\pm 0.18$.  This is still quite small
compared to the current atomic structure uncertainty 
($\pm 0.34$ in $Q_{{w}}$), but is 
as large as the uncertainty in $Q_{{w}}$ arising from
uncertainties in Standard Model radiative corrections (see
Fig.~\ref{cs_plot}).  All of the models we have considered predict
the charge radius in Cs within about 1$\%$, but the parameter
fits used to determine the Skyrme potentials are {\it based} in part
on observables, including charge radii, in nearby semi-magic even-even
nuclei.  There remain various possible sources of concern that a value
of $\delta R_{{n}}^2\approx \pm 1 {\rm fm}^2$ may still be an
underestimate. For example, $^{133}$Cs is a deformed, odd-Z
nucleus. Most nuclear structure calculations for large nuclei assume
spherical symmetry with at least partially closed nuclear
subshells. Pairing and blocking effects make calculations with odd N
or Z less reliable,\cite{Dobaczewski} as evidenced by the
failure of most Skyrme HF calculations to reproduce experimentally
observed ``even-odd" staggering of charge radius along isotope
chains.\cite{Atomic} In reference~\cite{Chen} pairing effects were
included, but deformation was included only in a semi-phenomenological
manner.

There exist other classes of nuclear structure models which give quite
different predictions for neutron properties, for example, relativistic
Hartree models based on a modified Walecka-model nuclear
Lagrangian.\cite{Relativistic} These models have seen significant
improvements in recent years, and may now be viewed as competitive with
more established Skyrme models in terms of their predictive power over
a wide variety of observables throughout the periodic table. In a
recent paper comparing models,~\cite{Patyk}
$R_{{n}}^2/R_{{p}}^2$ for $^{138}$Ba (the nearest
even-even semi-magic nucleus above Cs) ranged from 1.03 in a Skyrme
model to more than 1.08 in the relativistic models.  The difference in
predicted $R_{{n}}^2$ between these two models alone exceeds
1~fm$^2$.  For $^{136}$Xe, $R_{{n}}^2/R_{{p}}^2$ values
vary from 1.04 to 1.09, with the predicted $R_{{n}}^2$ differing
by well over 1 fm$^2$.  In another recent paper comparing
models,~\cite{Dobaczewski} the predictions for $R_{{n}}$ in
$^{124}$Sn (with a value of N/Z similar to $^{133}$Cs) varied by more
than 2~fm$^2$ between extreme models, a spread of over 8$\%$.  Again,
these calculations are primarily for even-even nuclei; relativistic
models have not yet been used to calculate self-consistently in the
neighboring unpaired (odd Z) cases.  This only {\it adds} to the
uncertainty in the prediction of a model spread for the case of Cs.
Based on these spreads, it appears that current nuclear theory
yields an uncertainty of at least 4 or 5\% in $R_n$.

%

The uncertainties in neutron distributions discussed so far
arise from disagreements between model predictions. It is important to
note that the neutron rms radius has never been directly measured in
any isotope of Cs.  Indeed, it is extremely difficult to measure
$R_{{n}}$ in any nucleus - the most accurate measurements of
charge radii come from electromagnetic interactions, which are
dominated by the proton distribution.  Elastic magnetic scattering is
affected mostly by unpaired (valence) nucleons, which does not allow
for a detailed or accurate measure of the bulk rms neutron radius. Data
from strong interaction probes measure the ``matter radius", but are
somewhat more sensitive to surface effects, and suffer from some poorly
controlled systematic theoretical uncertainties arising from the models
required in analyzing strong interaction observables. For example,
there exist data from polarized proton elastic scattering on heavy
nuclei.~\cite{Ray}  The data are statistically of high quality,
and are frequently viewed as an accurate experimental measure of
$R_{{n}}$ in several heavy nuclei, including Sn and Pb. However,
the systematic uncertainties in extracting $R_{{n}}$, including
choice of optical model and spurious variations in the result as a
function of experimental beam energy, easily approach 5$\%$ or
more.  Other data, including pion or alpha scattering, suffer from
similar uncertainties.  The experimentally extracted average value from
polarized proton scattering~\cite{Ray} and pionic
atoms~\cite{Oset} for $R_{{n}}$ in $^{208}$Pb differ by around
3$\%$.  

Even if strong interaction measurements can be argued to provide an
accurate measure of the neutron rms radius, the weak interaction is
sensitive to the spatial distribution of weak charge, which can not be
exactly identified with neutrons or protons, but also includes effects
of other nuclear degrees of freedom including e.g. meson exchange, and
is more sensitive to non-surface density variations.  A
parity violating electron scattering
experiment~\cite{Welliver,souder,musolf} could directly measure the
weak charge distribution, precisely what is needed for the
interpretation of atomic PNC as a standard model test, and would be of
clear value. Even if measured on another nucleus, the additional
constraint on nuclear models should increase
confidence in the predicted neutron distribution in Cs.  As can be
seen from Eq.~\ref{approxiii}, high precision atomic PNC measurements
on significantly higher Z nuclei are {\it more} sensitive to the
neutron distribution than in the case of Cs.  Thus, a measurement of
atomic PNC on extremely heavy nuclei might also be used as a measure
of the neutron distribution, which in turn could be used to constrain
the isovector parameters in the nuclear models, and thus increase the
reliability of the predictions for Cs.

To summarize, $\Delta Q_{{w}}^{{n-p}}$ is the deviation
between the experimentally extracted weak charge and Standard Model
predictions due solely to differences in neutron and proton weak charge
spatial distributions.  The predicted value is small, typically of
order $0.1$, but with an uncertainty larger than the value itself,
arising mostly from uncertainties in $R_n$.  This should be compared to
the nominal value of the weak charge, $Q_{{w}}^{\rm St.
Mod}=-73.20(13)$.  The effect of uncertain nuclear structure is thus
comparable to the present uncertainties involved in the Standard Model
prediction, and for $R_{{n}} > R_{{p}}$ is of the right
sign to partially explain the experimental discrepancy in Cs. For these
reasons, it should be included in any future atomic PNC tests of the
Standard Model.  With any significant further reduction in the
uncertainties in atomic theory calculations, this nuclear contribution
may eventually limit the level at which Standard Model tests can be
performed with atomic PNC on Cs.  To reliably reduce this uncertainty
would require additional direct experimental input on neutron
distributions, most likely from parity violating electron scattering at
low momentum transfer, e.g. at a facility such as Jefferson Lab.


\acknowledgments

We are grateful to Jacek Dobaczewski and J. Decharge for providing
neutron and proton densities in Cs, calculated with Skyrme and Gogny
potentials.  Our work was supported in part under U.S. Department of
Energy contract \#DE-FG03-93ER40774.


\begin{thebibliography}{99}

\bibitem{Wieman99} S.~C.~Bennett and C.~E.~Wieman,
Phys. Rev. Lett.{\bf 82}, 2484 (1999)

\bibitem{Wieman98} C.~S.~Wood  et al, 
Science {\bf 275}, 1759 (1997).

\bibitem{Marciano} W.~Marciano and J.~Rosner, Phys. Rev. Lett. {\bf 65},
2963 (1990); {\bf 68}, 898(E) (1992)

\bibitem{Johnson} S.~A.~Blundell, J. Sapirstein, W.~R.~Johnson, Phys. Rev. {\bf
D45} 1602 (1992)

\bibitem{Dzuba} V.~A.~Dzuba et al, Phys Lett. {\bf A 141}, 147 (1989)

\bibitem{Fortson} E.~N.~Fortson et al,
Phys. Rev. Lett. {\bf 65}, 2857 (1990),
S.~J.~Pollock et al,
Phys. Rev. C {\bf 46}, 2587 (1992)


\bibitem{Chen} P.~Q.~Chen and P.~Vogel, Phys. Rev. {\bf C 48} (1993) 1392



\bibitem{Brack} M.~Brack, C.~Guet, H.-B.~Hakansson, 
Phys. Rep {\bf 123}, 275 (1985)

\bibitem{Skyrme} 
M.~Beiner et al, Nucl. Phys {\bf A238}, 29 (1975),
J.~Bartel et al, Nucl. Phys {\bf A386}, 70 (1982),
P.~Honche et al, Nucl. Phys {\bf A443}, 39 (1985) 

\bibitem{Relativistic} 
B.~Serot, J.~D.~Walecka, Adv. Nucl. Phys {\bf 16} 1 (1986),  
P.-G.~Reinhard, Rep. Prog. Phys {\bf 52} 439 (1989),
G.~Lalazissis et al, Phys. Rev. {\bf C55} 540 (1997)

\bibitem{Patyk} Z.~Patyk et al, Phys. Rev. {\bf C59} 704 (1999)

\bibitem{Atomic}  E.~G.~Nadjakov et al, 
At. Data Nucl. Data Tables {\bf 56}, 133 (1994)

\bibitem{Ray} L.~Ray and G.~W.~Hoffman, Phys. Rev. {\bf C 31}, 538 (1985),
C.~J.~Batty et al, Adv. Nucl Phys. {\bf 19}, 1 (1989)


\bibitem{Welliver} S.~J. Pollock and M.~Welliver, CU preprint NPL-1164

\bibitem{souder} R.~Michaels and P.~A.~Souder, 
Proposal E99-102 to PAC-15, Jefferson Lab.

\bibitem{musolf} M.~J.~Musolf et al,
Phys. Rep. {\bf 239}, 1 (1994).

\bibitem{Dobaczewski} J.~Dobaczewski et al, Z. 
fur Phys. {\bf A 354}, 27, (1996)
J.~Dobaczewski et. al, Phys. Rev. {\bf C53} 2809 (1996), 
nucl-th/9901036. 


\bibitem{dobiii} 
J.~Dobaczewski and J.~Decharge, private communications. 



\bibitem{Oset} C.~Garcia-Recio, J.~Nieves, E.~Oset, Nucl. Phys. {\bf A547},
473 (1992)

\end{thebibliography}
\end{document}